\documentclass[x11names,11pt]{article}
\usepackage[paper=a4paper,margin=1in]{geometry}

\usepackage{jheppub} 
\usepackage{setspace}

\pdfoutput=1

\usepackage[utf8]{inputenc}
\usepackage{uniinput}
\usepackage{microtype} %

\usepackage{mathtools,cleveref,xcolor,enumitem, subcaption}
\definecolor{dark-green}{rgb}{0.1,0.4,0}
\definecolor{NiceBlue}{rgb}{0.30196,0.55294,0.57647}

\usepackage{hyperref}
\hypersetup{
      colorlinks=true,
      linkcolor=dark-green,
      citecolor=Red4,
      urlcolor=dark-green,
      linktoc=page
}
\newcommand{\bea}{\begin{eqnarray}}
\newcommand{\eea}{\end{eqnarray}}

\def\Im{ \hbox{\rm Im}}

\def\bra{\langle}\def\ket{\rangle}

\def\Re{\hbox {\rm Re}}
\def\Im{\hbox {\rm Im}}

\begin{document} 
\begin{flushright}
  \footnotesize UUITP-17/26\\
  \normalsize
  \end{flushright}
\thispagestyle{empty}

\vspace*{1.5cm}
\begin{center}

\begin{doublespace}
{\bf {\Large Quantum nucleation of black hole mimickers\\ via chaos dominated tunneling
}}
\end{doublespace}

\begin{center}

\vspace{1cm}

\hypersetup{urlcolor=black}
{\bf \href{mailto:ulf.danielsson@physics.uu.se}{Ulf Danielsson}}$^{1}$, {\bf \href{mailto:suvendu.giri@physics.uu.se}{Suvendu Giri}}$^{1}$, {\bf \href{mailto:vyshnav.vijay.mohan@gmail.com}{Vyshnav Mohan}}$^{2}$, \textbf{and} {\bf \href{mailto:lth@hi.is}{Larus Thorlacius}}$^{2}$

\bigskip

$^{1}$\,Institutionen f\"or fysik och astronomi, Uppsala Universitet\\
Box 803, SE-751 08 Uppsala, Sweden

\bigskip

$^{2}$\,Science Institute, University of Iceland\\
Dunhaga 3, 107 Reykjav\'ik, Iceland

  \end{center}

\vspace{1.5cm}
{\bf Abstract}
\end{center}
\begin{quotation}
\noindent
Black hole mimickers, ultracompact horizonless objects, have been proposed as alternatives to black holes in a variety of settings, including extensions of general relativity and scenarios involving matter sectors beyond the Standard Model. Their formation in gravitational collapse of matter requires quantum mechanical tunneling to occur on a length scale of the order of the Schwarzschild radius of the corresponding black hole. We propose a mechanism, based on multichannel enhancement catalyzed by quantum chaotic dynamics, that can dramatically amplify the quantum mechanical transmission across a tunneling barrier. We explore, in particular, how the nucleation of a string theoretic black shell is enhanced via this mechanism. We anticipate similar results to hold for other proposed black hole mimickers.

\end{quotation}

\setcounter{page}{0}
\setcounter{tocdepth}{2}
\setcounter{footnote}{0}
\newpage

\parskip 0.1in
 
\setcounter{page}{2}

\setcounter{tocdepth}{1}

{\hypersetup{linkcolor=black}
\tableofcontents
}

\section{Introduction}

For black hole mimickers, i.e. horizonless ultracompact objects,\footnote{For a review of various proposed black hole mimickers, see \cite{Bambi:2025wjx,Cardoso:2019rvt}.} to be a viable alternative to black holes there needs to exist an efficient mechanism for their formation. Typically, one needs a tunneling event, with a probability that is enhanced thanks to the large entropy of the final state. This has been argued for in the context of fuzzballs, \cite{Mathur:2008kg,Kraus:2015zda}, frozen stars, \cite{Brustein:2025yvf}, as well as in the context of AdS black shells, \cite{Danielsson:2017riq}.
The argument is that the huge number of states that become available when a black mimicker is formed compensates for the suppression caused by the tunneling. However, it is far from clear that just having a large number of final states actually increases the tunneling rate. A more relevant criterion is whether there are a large number of {\it channels} available for which the transition can proceed without exponential suppression. 

Let us imagine an excited discrete state (such as an atomic state) that decays to a continuum (e.g. electromagnetic radiation). The probability for this to happen is given by Fermi's golden rule. Now put the atom at the center of a large isolating shell with a macroscopic number of degrees of freedom at low energy. The shell will absorb the light, and distribute the energy among its many degrees of freedom. The number of possible states is huge and so is its entropy. Will this enhance the decay rate of the atom? Clearly not. It will still be given by the golden rule. If the shell is small, and the boundary conditions of the electromagnetic radiation are changed, the rate may change a bit, but this has nothing to do with the entropy of the final state of the shell.

We could try to improve the treatment by considering the transition from the excited atom to the thermalized shell in a single go. Perhaps the full quantum treatment of the entire process would change the conclusion? Actually, it will not. The final states given by the golden rule are orthogonal and there are no quantum interference terms to worry about. Hence, there is no difference between considering the full process in one step, or making a cut to a two step process. It is easy to see why. Let us denote the coupling in the golden rule by $g$. It is true that all the final states of the shell are indirectly coupled to the decaying atom, but the coupling is spread out over all the $N$ degrees of freedom of the shell. So, even if you would expect an enhancement from the number of final states by $N$, the effective coupling to each of them is reduced to $g/\sqrt{N}$. In total we find $g^2 \rightarrow N \times(g/\sqrt{N})^2= g^2$, that is, no change. The decay to radiation is a bottle neck, determined by the small value of the coupling $g$.

To get an actual enhancement we need not just a large number of final states but a large number of available channels.  In \cref{sec:enhancementsection}, we construct a large class of quantum mechanical models where enhanced tunneling is explicitly realized. Concretely, we study a non-relativistic particle tunneling across a potential barrier. The particle carries $N$ internal degrees of freedom that interact under the barrier. The dynamics of the particle is governed by a Schrödinger equation in which the potential is the sum of the barrier potential and an interaction matrix encoding couplings between the internal states. In the large $N$ limit, we show that whenever this potential matrix has $O(N)$ negative eigenvalues, the tunneling amplitude is no longer exponentially suppressed.

We find the strongest enhancement when the couplings between the large number of internal states is given by Gaussian random matrices, which are well known to capture the dynamics of \emph{chaotic} interactions. Chaotic interactions of this type are ubiquitous in large many-body systems and are expected to also arise in black shells. For a shell to be ``black'', information carried by infalling matter must be scrambled rapidly, and such rapid scrambling is a hallmark of maximally chaotic dynamics. 
A considerably weaker enhancement mechanism referred to as chaos assisted tunneling (CAT), where a tunneling barrier is effectively lowered by chaotic interactions, is known from other branches of physics \cite{DASSO1983381,e26020144}. We instead refer to the phenomenon described below, where chaotic interactions between an enormous number of internal states open up a sufficient number of channels to actually overcome the exponential suppression of the original tunneling barrier, as \emph{chaos dominated tunneling} (CDT).

The paper is organized as follows. In \cref{sec:enhancementsection}, we study the quantum mechanical scattering of a one-dimensional particle, with a large number of internal states, incident on a potential barrier and present examples where transmission across the barrier is significantly enhanced by interactions between particle states. In \cref{sec:blackshellsection}, we argue that the resulting phenomenon of chaos dominated tunneling can strongly enhance the probability of black shell nucleation in gravitational collapse of matter. \Cref{sec:localitysection} discusses how the apparent breakdown of spacetime locality when a black shell nucleates via an inherently stochastic process of quantum mechanical tunneling is resolved. A brief discussion of our results and possible next steps follows in \cref{sec:discussionsection}.

\section{Multichannel enhancement}
\label{sec:enhancementsection}

Tunneling can be enhanced if there are multiple channels available to the system that is making the transition. A necessary condition is the presence of several internal states that interact during the tunneling process. As a consequence, a complex object may tunnel more easily than a simple one, provided that its degrees of freedom remain coherent. An early example of this phenomenon came from the fusion of heavy ions, where it was noted that nuclei with more internal degrees of freedom tunneled more rapidly than you would naively expect \cite{DASSO1983381}. In this section, we provide a theoretical \textit{proof of principle} for multichannel enhancement by studying quantum mechanical tunneling in a sequence of one-dimensional systems of increasing complexity, where the effect of having a large number of interacting states can be explicitly worked out.

\subsection{Tunneling through a square potential barrier}

We start with a simple example where a non-relativistic particle tunnels trough a square barrier of height $V_0$ and width $a$. To emulate the large number of degrees of freedom carried by a black shell, we assume that the particle has $N$ internal states that interact with each other under the barrier but propagate freely otherwise. 
The time-independent Schrödinger equation for $\Psi=\left(\psi_1,\psi_2,...,\psi_N\right)_T$ is
\begin{equation}
    -\frac{1}{2m}\frac{d^2}{dx^2}\Psi+ \Theta(x)\Theta(a-x) \mathcal{M} \Psi=E \Psi \,,
    \label{schrodingereq0}
\end{equation}
where $\Theta(x)$ is the Heaviside step function and the matrix valued potential is given by
\begin{equation}
  \renewcommand{\arraystretch}{1.3}
  \setlength{\arraycolsep}{6pt}
  \mathcal{M}=\left[\begin{matrix}V_{0}&-g&-g&\dots &-g&-g\\ -g&V_{0}&-g&\dots &-g&-g\\ -g&-g&V_{0}&\dots &-g&-g\\ \vdots &\vdots &\vdots &\ddots &\vdots &\vdots \\ -g&-g &-g&\dots &V_{0}&-g\\ -g&-g&-g&\dots &-g&V_{0}\end{matrix}\right].\label{eq:mmatrix1}
  \renewcommand{\arraystretch}{1.0}
\end{equation}
The diagonal entries reflect the barrier height, which is the same for all the internal states, and, for now, we take the pairwise coupling $-g$ to be the same between all the internal states.

We are interested in the probability for low-energy particles, with $0<E\ll V_0$, to tunnel through the potential barrier. The scattering problem separates into $N$ individual problems for the eigenvectors of $\mathcal{M}$, where each eigenvector faces a barrier height given by the corresponding eigenvalue. 
It is easy to see that the matrix $\mathcal{M}$ in \cref{eq:mmatrix1} has $N-1$ degenerate eigenvectors with eigenvalue $\lambda =V_0+g$ and a single eigenvector, $\Psi_{sc} =\frac{1}{\sqrt{N}}(1,1,1,...,1)$, with a lower eigenvalue $\lambda = V_0-(N-1)g$. 

To be explicit, the transmission coefficient for a particle with energy $E$ through a square barrier of height $V_0$ and width $a$ is given by
\begin{equation}
    T=\left[1+\frac{(k^2+\kappa^2)^2}{4k^2\kappa^2} \sinh^2\kappa a \right]^{-1},
\end{equation}
where $k^2 = 2m E$, and $\kappa^2 =2m(V_0-E)$. The transmisson probability is exponentially suppressed for $E\ll V_0$.
For a small coupling constant, $0<g\ll V_0$, the barrier height for the degenerate eigenvectors remains close to $V_0$ and their tunneling rate remains exponentially suppressed. However, for the $\Psi_{sc}$ eigenvector, the barrier is lowered:  $V_0\rightarrow V_0-(N-1)g$. It represents a \textit{bright channel} that will dominate the tunneling for large values of $N$. All the other modes decay more rapidly and become dark. In fact, if $(N-1)g> V_0$ the exponential suppression goes away completely!

We see that a physical condition required for the enhancement is that the object that tunnels has a large number of internal states $N\gtrsim V_0/g$. Furthermore, the object needs to be in a coherent superposition of all these states as a macroscopic quantum object.
There is a further twist that we need to take into account. While we have removed the heavy suppression of the exponential, the tunneling probability is still suppressed due to the projection of the in-state onto the single bright channel  $\Psi_{sc}$ through
\begin{equation}
\bra\Psi_{in}|\Psi_{sc} \ket^2=\frac{1}{N}\,.
\end{equation}
In order to overcome this remaining suppression factor, one needs to increase the number of available bright channels. Below, we work out  concrete cases where this is achieved.

\subsection{Towards a bright future}

Our next example improves on the previous section in two ways. First of all, we replace the square barrier with a smooth Pöschl-Teller potential barrier
to avoid reflections from sharp edges,
\begin{equation}
    -\frac{1}{2m}\frac{d^2\Psi}{dx^2}+ \frac{1}{\cosh^2 \left(x/a\right)}\mathcal{M} \Psi=E \Psi \,.\label{Schrodingereq}
\end{equation}
The transmission coefficient for an eigenvector of $\mathcal{M}$ that corresponds to a large positive eigenvalue $\lambda_n$, satisfying $ma^2\lambda_n\gg 1$, is then given by
\begin{equation}
   T_n=\frac{\sinh^2\pi ka}{\sinh^2\pi ka+\cosh^2\left(\frac{\pi}{2}\sqrt{8ma^2\lambda_n-1}\right)}\,,
\end{equation}
where $k^2 = 2m E$. It immediately follows that $T_n$ is exponentially suppressed when $E\ll \lambda_n$. If the $\mathcal{M}$ eigenvalue is negative, the potential barrier is replaced by a Pöschl-Teller potential well, and the transmission probability is no longer exponentially suppressed.
The same is true for small positive eigenvalues, for which the potential barrier is too low to significantly suppress the transmission probability. 

The second improvement is to consider more general coupling matrices to increase the number of available bright channels. An interesting and tractable class is provided by Hermitian circulant matrices,
\begin{equation}
\renewcommand{\arraystretch}{1.3}
\setlength{\arraycolsep}{6pt}
\mathcal{M}= \begin{bmatrix}
c_0 & c_{N-1} & \dots & c_2 & c_1 \\
c_1 & c_0 & c_{N-1} & \dots & c_2 \\
\vdots & c_1 & c_0 & \ddots & \vdots \\
c_{N-2} & \dots & \ddots & \ddots & c_{N-1} \\
c_{N-1} & c_{N-2} & \dots & c_1 & c_0
\end{bmatrix}\,,
\label{mmatrix}
\renewcommand{\arraystretch}{1}
\end{equation}
where the Hermiticity condition requires $c_{N-k} = c_k^*$. For simplicity, we assume that $N$ is odd. The eigenvalues of the circulant matrix are given in closed form by
\begin{equation}
    \lambda_n = \sum_{k=0}^{N-1} c_k \,\omega^{nk}\,, 
    \qquad \omega = e^{2\pi i/N}\,.
\end{equation}
With $\theta_n = 2\pi n/N$, and the identification $c_{-k}\equiv c_{N-k}$ for $k\in\{1,\ldots,(N-1)/2\}$, the eigenvalues in the $N\to\infty$ limit are 
given by \cite{Toeplitzreview},
\begin{equation}
   \lambda(\theta) = \sum_{k=-\infty}^{\infty} c_k\, e^{ik\theta}\,, 
   \qquad \theta \in (-\pi,\pi]\,.
   \label{eq:lambdasum}
\end{equation}
The condition for enhanced tunneling is simply  that $\lambda(\theta) \lesssim 0$ for some range of $\theta \in (-\pi,\pi]$.

As a concrete example, take $c_0 = V_0$ and $c_{k} = -g$ for $|k|\leq M$, with all remaining coefficients vanishing. Then the sum in \cref{eq:lambdasum} terminates at $k=M$ and is easily evaluated to give,
\begin{equation}
    \lambda(\theta) = V_0 -g \frac{\sin( (M+1/2)\theta)}{\sin (\theta/2)}\,.
\end{equation}
When $(2M+1)g > V_0$, the eigenvalues near $\theta=0$ are negative. The range of negative eigenvalues, $\vert\theta\vert<\theta_0$, depends on the ratio $V_0/g$ and on $M$. In the limit $M\gg V_0/g$ one finds 
\begin{equation}
    \theta_0 \approx \frac{\sqrt{6}}{M}\,.
\end{equation}
but the precise value of $\theta_0$ is not important for our argument. As long as a finite fraction the eigenvalues are negative, there will be an $O(N)$ number of bright channels.
Thus we have found an example of a system where tunneling is no longer significantly suppressed, and the enhancement is due to the presence of a large number of coupled degrees of freedom. This is exactly what needs to happen for a successful black hole mimicker.

\subsubsection{Wigner Random Matrices} 

For our final example we consider a coupling matrix of the form
\begin{equation}
\mathcal{M} = V_0 \mathbb{I} - g \mathcal{J}\,,
\end{equation}
where $\mathbb{I}$ is the $N\times N$ unit matrix and $\mathcal{J}$ is a real symmetric (or Hermitian) random matrix with 
zero mean and variance $\sigma$ per entry. By Wigner's semicircle law \cite{Wigner1951}, 
the limiting eigenvalue density of $\mathcal{J}$ is supported on 
$[-2\sigma \sqrt{N}, 2\sigma\sqrt{N}]$, so the spectrum of $\mathcal{M}$ is supported on 
$[V_0 - 2\sigma g\sqrt{N},\, V_0 + 2\sigma g\sqrt{N}]$. Negative eigenvalues of $\mathcal{M}$ 
therefore exist, and constitute an $O(N)$ fraction, whenever
\begin{equation}
    2\sigma g \sqrt{N} > V_0\,.
\end{equation}
Random matrix theory captures the spectral statistics of quantum chaotic systems \cite{Bohigas:1983er} and the enhancement mechanism we are describing is intimately connected to so-called chaos assisted tunneling (CAT).  Experimental realizations include tunneling in micro-wave and optical cavities, as well as cold atoms trapped in electromagnetic fields, for a recent review see \cite{e26020144}. In these systems, the tunneling barrier is effectively lowered due to  chaotic interactions and this leads to enhanced transition rates. In the case at hand, however, the effect is much stronger. Due to the enormous number of states in the gravitational system, the random interactions not only lower the barrier, but overwhelm it, for a finite fraction of the eigenvectors of $\mathcal{M}$,  leading to an $O(N)$ number of bright channels for which the transition is no longer exponentially suppressed. We refer to this as \textit{chaos dominated tunneling} (CDT).  As we will argue below, this mechanism is the most promising to be at work in a black hole mimicker.

\section{Application to the AdS black shell}
\label{sec:blackshellsection}

The chaos dominated tunneling discussed in the previous section can have important consequences for macroscopic tunneling processes. A particularly interesting example is the formation of a black hole mimicker from collapsing matter. While black hole mimickers have been investigated for decades, an outstanding question is how collapsing matter can form a mimicker instead of a black hole. Some classical proposals have been made, but they often rely on the existence of matter with no known physical origin (see e.g. \cite{Jampolski:2025lrh}). As a result, a purely classical formation mechanism that avoids black hole formation remains an open question.

A much more plausible possibility is a quantum mechanical formation mechanism, which has been argued for several types of black hole mimickers that have clear motivations from quantum gravity; see e.g. \cite{Kraus:2015zda,Danielsson:2017riq,Brustein:2025yvf}. In these scenarios, the collapsing matter tunnels quantum mechanically into the mimicker configuration instead of forming a black hole. The open question here is to understand how the exponentially suppressed tunneling probability can be overcome, so that the nucleation of a mimicker becomes sufficiently likely.

In this section, we will focus on one particular black hole mimicker, the AdS black shell \cite{Danielsson:2017riq}, and argue that chaos dominated tunneling can greatly enhance the probability of a collapsing shell of matter to tunnel quantum mechanically  into a black shell.

\subsection{Review of a nucleating black shell}

The AdS black shell model uses ingredients from string theory to propose that our 4D Minkowski universe is metastable, with a more stable AdS$_4$ vacuum (characterized by a microscopic AdS length $L \coloneqq 1/k$) that can nucleate locally within our universe. Such a nucleation replaces a region of our 4D Minkowski space with a spherical bubble of AdS$_4$ vacuum enclosed by a thin shell made of string-theoretic branes, together with radiation living on the shell. 
Crucially, although the construction is motivated by string theory, the \emph{resulting dynamics is fully described within 4D general relativity}, with the string-theoretic input entering only through the effective matter content of the shell, which admits a natural interpretation in terms of branes and associated string states on the world-volume of the branes. 
The shell consists of a surface tension component (with equation of state $p=-ρ$), and radiation (with equation of state $p=ρ/2$, at subleading order in $k$). This remarkably simple matter model mimics a black hole surprisingly well, and has been shown to possess non-linear dynamical stability under radial perturbations \cite{Danielsson:2021ykm}.\footnote{To achieve stability, it is crucial that the matter components can interact and redistribute their energy.} Rotating black shells that mimic Kerr black holes, together with concrete predictions for quadrupole moments that differ from Kerr, were studied in \cite{Danielsson:2023onu}. Electromagnetic properties of the black shell have been studied in \cite{Giri:2024cks}, where it is modeled as an effective dielectric membrane with large electric and magnetic permittivity (and conductivity), with implications for accretion physics and jet formation. Gravitational-wave and radio VLBI observables were also investigated in the same work. 

To understand the essence of the quantum mechanical nucleation process, let us focus on a spherical AdS black shell. A convenient way to describe the dynamics is to formulate the black shell as a thin shell is GR, with its radius $r(t)$ as the dynamical degree of freedom. We can write the metric on the brane as (making a gauge choice of the lapse function being trivial)
\begin{equation}\label{eq:brane-metric}
    ds²_Σ = -dt² + r(t)² dΩ²₂\,,
\end{equation}
and that of the 4D bulk inside and outside the shell as
\begin{equation}
    ds_\pm² = -f_\pm(r) dT_\pm² + \frac{dr²}{f_\pm(r)} + r² dΩ₂²\,,
\end{equation}
with AdS$_4$ inside: $f_-(r) = 1 + k²r²$ and Schwarzschild outside: $f_+(r)=1-2G_4M/r$.
From the Einstein-Hilbert action and Israel's junction conditions, we can write an effective Lagrangian for this radial degree of freedom, which can then be used to write a Hamiltonian constraint \cite{Ansoldi:1997hz}:
\begin{equation}\label{eq:Hamiltonian-constraint}
    H=4\pi r^2 \sigma_2-\frac{r}{G_4}\left[\left(\sqrt{f_-(r)}-\sqrt{f_+(r)}\right)^2+2\sqrt{f_-(r)f_+(r)}\left( 1-\cosh \frac{G_4p}{r}\right)\right]^{1/2}
\end{equation}
with $H=0$. Here $p$ is the momentum conjugate to $\dot{r}\coloneqq d r(t)/dt$.
We can quantize this Hamiltonian with
\begin{equation}
    \hat{p} = - i \hbar \frac{1}{r} \frac{d}{dr}(r\, \cdot)\,,\qquad
    \hat{r} = r\,,
\end{equation}
which satisfy the canonical commutation relation $[\hat{r},\hat{p}]=i\hbar$. The Hamiltonian constraint $\hat{H}|Ψ\rangle = 0$ then selects the physical states from the full Hilbert space.

Interestingly, this Hamiltonian is not quadratic in momentum $p$, but rather has a non-linear dependence on it with a $\cosh(G_4\hat{p}/r)$. Moreover, it also has a square root, reminiscent of its relativistic origins.
In analogy with how the Cartesian momentum operator generates an infinitesimal translation $e^{ϵ ∂_x} f(x) \approx f(x+ϵ)$, the Hamiltonian contains the operator
\begin{equation}
    \cosh{\frac{G_4 \hat{p}}{r}} \approx \exp \left[\frac{\hbar }{r²}∂_r (r\, \cdot)\right]\,,
\end{equation}
which generates translations of the area:
\begin{equation}
    \exp\left[\frac{\hbar}{r²}∂_r (r\, \cdot)\right] r² \approx r² + 3\hbar G_4 \approx r² + 3\ell_p²\,.
\end{equation}
This implies that area, proportional to $\sim r^2$, is quantized in units of $G_4$, i.e. in units of the Planck length squared.

If $p$ is small, $1-\cosh(p) \approx p²/2$, and we recover a Schrödinger-like equation that is quadratic in the momentum.\footnote{In \cite{Danielsson:2021tyb} the same reasoning was applied in 5D to describe the nucleation of a universe with a positive cosmological constant in the dark bubble model. It was shown to agree with the Wheelet--DeWitt equation of quantum cosmology in Vilenkin's formulation.}
For general momentum, however, the exponentials in the Hamiltonian generate difference operators, giving rise to a difference equation instead of a differential equation. Such equations are common in condensed matter physics, where they describe electrons moving on a crystal lattice, with the lattice spacing introducing the discreteness (see e.g. \cite{diff-eqs}).

To study the tunneling process, and to make the discussion less dependent on the detailed construction of the black shell and its stabilization mechanism, we assume that the infalling matter is itself in the form of a thin shell. Once it enters within its own Buchdahl radius, a black shell nucleates, capturing the infalling matter and converting it into degrees of freedom living on the shell. Consequently, the tunneling takes place entirely in flat space without encountering any matter.

\begin{figure*}[t]
    \centering
    \begin{subfigure}[t]{0.48\textwidth}
        \centering
        \def\svgwidth{\linewidth}
        \begingroup%
  \makeatletter%
  \providecommand\color[2][]{%
    \errmessage{(Inkscape) Color is used for the text in Inkscape, but the package 'color.sty' is not loaded}%
    \renewcommand\color[2][]{}%
  }%
  \providecommand\transparent[1]{%
    \errmessage{(Inkscape) Transparency is used (non-zero) for the text in Inkscape, but the package 'transparent.sty' is not loaded}%
    \renewcommand\transparent[1]{}%
  }%
  \providecommand\rotatebox[2]{#2}%
  \newcommand*\fsize{\dimexpr\f@size pt\relax}%
  \newcommand*\lineheight[1]{\fontsize{\fsize}{#1\fsize}\selectfont}%
  \ifx\svgwidth\undefined%
    \setlength{\unitlength}{424.55996704bp}%
    \ifx\svgscale\undefined%
      \relax%
    \else%
      \setlength{\unitlength}{\unitlength * \real{\svgscale}}%
    \fi%
  \else%
    \setlength{\unitlength}{\svgwidth}%
  \fi%
  \global\let\svgwidth\undefined%
  \global\let\svgscale\undefined%
  \makeatother%
  \begin{picture}(1,0.57760974)%
    \lineheight{1}%
    \setlength\tabcolsep{0pt}%
    \put(0,0){\includegraphics[width=\unitlength,page=1]{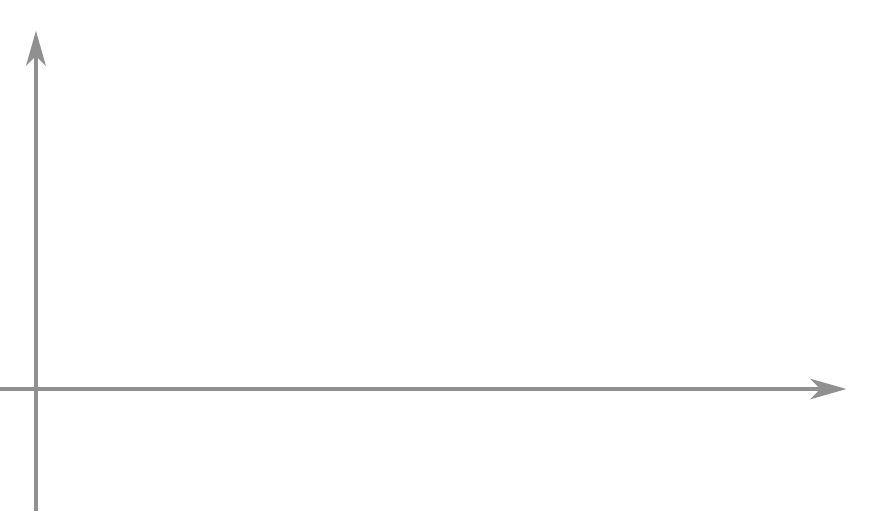}}%
    \put(0.01943188,0.55999156){\color[rgb]{0,0,0}\makebox(0,0)[lt]{\lineheight{1.25}\smash{\begin{tabular}[t]{l}$V(r)$\end{tabular}}}}%
    \put(0.96982766,0.13249012){\color[rgb]{0,0,0}\makebox(0,0)[lt]{\lineheight{1.25}\smash{\begin{tabular}[t]{l}$r$\end{tabular}}}}%
    \put(0.78610803,0.08302714){\color[rgb]{0,0,0}\makebox(0,0)[lt]{\lineheight{1.25}\smash{\begin{tabular}[t]{l}$r_0$\end{tabular}}}}%
    \put(0,0){\includegraphics[width=\unitlength,page=2]{pot1.pdf}}%
    \put(0.21905033,0.08302714){\color[rgb]{0,0,0}\makebox(0,0)[lt]{\lineheight{1.25}\smash{\begin{tabular}[t]{l}$R$\end{tabular}}}}%
  \end{picture}%
\endgroup%

        \caption{The effective potential (blue) terminates when the tunneling shell encounters the collapsing matter shell (red vertical line) at $r=R$.}
        \label{fig:potential-a}
    \end{subfigure}
    \hfill
    \begin{subfigure}[t]{0.48\textwidth}
        \centering
        \def\svgwidth{\linewidth}
        \begingroup%
  \makeatletter%
  \providecommand\color[2][]{%
    \errmessage{(Inkscape) Color is used for the text in Inkscape, but the package 'color.sty' is not loaded}%
    \renewcommand\color[2][]{}%
  }%
  \providecommand\transparent[1]{%
    \errmessage{(Inkscape) Transparency is used (non-zero) for the text in Inkscape, but the package 'transparent.sty' is not loaded}%
    \renewcommand\transparent[1]{}%
  }%
  \providecommand\rotatebox[2]{#2}%
  \newcommand*\fsize{\dimexpr\f@size pt\relax}%
  \newcommand*\lineheight[1]{\fontsize{\fsize}{#1\fsize}\selectfont}%
  \ifx\svgwidth\undefined%
    \setlength{\unitlength}{424.55996704bp}%
    \ifx\svgscale\undefined%
      \relax%
    \else%
      \setlength{\unitlength}{\unitlength * \real{\svgscale}}%
    \fi%
  \else%
    \setlength{\unitlength}{\svgwidth}%
  \fi%
  \global\let\svgwidth\undefined%
  \global\let\svgscale\undefined%
  \makeatother%
  \begin{picture}(1,0.57760974)%
    \lineheight{1}%
    \setlength\tabcolsep{0pt}%
    \put(0,0){\includegraphics[width=\unitlength,page=1]{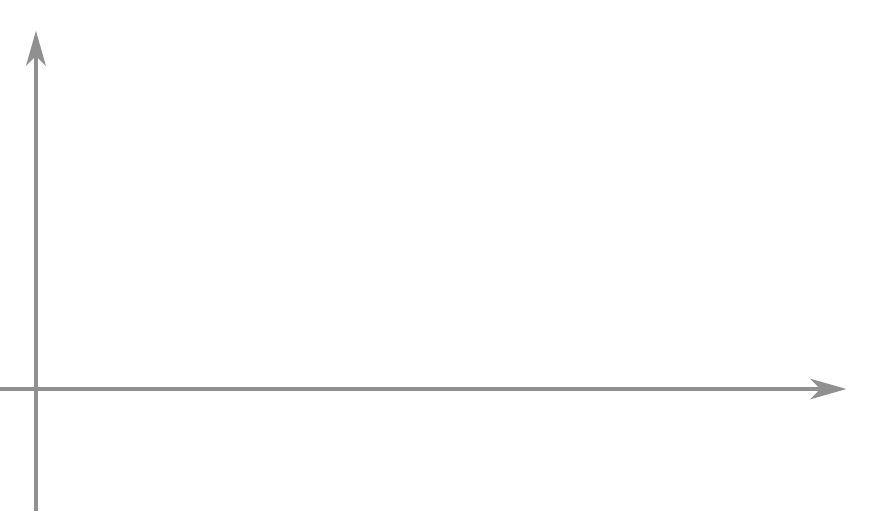}}%
    \put(0.01943188,0.55999156){\color[rgb]{0,0,0}\makebox(0,0)[lt]{\lineheight{1.25}\smash{\begin{tabular}[t]{l}$V(r)$\end{tabular}}}}%
    \put(0.96982766,0.13249012){\color[rgb]{0,0,0}\makebox(0,0)[lt]{\lineheight{1.25}\smash{\begin{tabular}[t]{l}$r$\end{tabular}}}}%
    \put(0,0){\includegraphics[width=\unitlength,page=2]{pot2.pdf}}%
    \put(0.5264274,0.0847465){\color[rgb]{0,0,0}\makebox(0,0)[lt]{\lineheight{1.25}\smash{\begin{tabular}[t]{l}$R$\end{tabular}}}}%
    \put(0,0){\includegraphics[width=\unitlength,page=3]{pot2.pdf}}%
  \end{picture}%
\endgroup%

        \caption{Schematic of effective potential for a finite-thickness collapsing matter distribution (red shaded region), interpolating between AdS-Minkowski and AdS-Schwarzschild regimes.}
        \label{fig:potential-b}
    \end{subfigure}
    \caption{Effective potential governing the radial motion of a nucleating black shell. (a) In the absence of collapsing matter, the AdS-Minkowski junction conditions give the full effective potential of \cref{eq:potential} (blue solid and dotted curves), with a classically forbidden region extending to the turning point $r_0$. In the thin-shell approximation for the collapsing matter (red dotted line), the AdS-Minkowski potential is valid only up to the intersection radius $R$, beyond which it is abruptly truncated and replaced by the AdS-Schwarzschild junction condition. (b) For a realistic collapsing matter distribution of finite thickness, the sharp cutoff is replaced by a smooth transition to the AdS-Schwarzschild potential, terminating at the turning point of the full potential $R$.}
    \label{fig:potential}
\end{figure*}

Let us begin with studying the formation of a bubble of AdS in flat space, without any infalling matter. Using 
$f_-(r) = 1+k²r²$ and $f_+(r)=1$,
the Hamiltonian constraint becomes
\begin{equation}\label{eq:Hamiltonian-flat}
H=4\pi r^2\sigma -\frac{r}{G_4}\left[2 + k²r² -2 \sqrt{1+k²r²}\cosh \frac{G_4p}{r}\right]^{1/2} \,.
\end{equation}
To understand the physics of the problem, let us look at the Hamiltonian in the static case, $p=0$, where it represents the total “potential” energy of the system. This potential 
\begin{equation}\label{eq:potential}
    V(r) = \frac{1}{G_4} \left(r-\frac{r²}{r_0} \right)\,,
\end{equation}
plotted in \cref{fig:potential-a} contains a classically forbidden region from $r=0$ to $r=r_0$, and an unbounded classically allowed region  for $r>r_0$. The shell must tunnel through the forbidden region before emerging at $r=r_0$. For a zero energy shell to nucleate at $r=r_0$, its tension is obtained by solving $H=0$:
\begin{equation}
    σ_0 
    \approx \frac{k}{4πG_4} - \frac{1}{4πG_4 r_0} + \mathcal{O}\left( 1/k r \right)\,.
\end{equation}
Here and throughout this paper, we neglect terms that are subleading in $1/(kr)$. We note that the size of the nucleated bubble, $r_0$, is directly related to the tension of the bubble. For nucleation the tension needs to be less than the critical value, and the closer it is, the larger the radius. To compute the quantum mechanical tunneling amplitude for tunneling from $r=0$ to $r=r_0$, we need the Euclidean momentum under the barrier ($p→i p_{\textrm{\scshape e}}$) which is obtained by solving $H(σ=σ_0)=0$:
\begin{equation}
    p_{\textrm{\scshape e}} \approx \frac{r \arccos(r/r_0)}{G_4}\,.
\end{equation}
The nucleation rate of a bubble in empty space needs to be extremely small, otherwise the universe would have made such a transition long ago. Thus, the radius $r_0$ needs to be very large and the tension close to its critical value.

So far we have analyzed the tunneling process in a region where the nucleating bubble evolves in an effectively flat background, well inside the collapsing matter shell. In this regime, the dynamics are governed by the Hamiltonian constraint in \cref{eq:Hamiltonian-flat}, which leads to a well-defined Euclidean momentum $p_{\textrm{\scshape e}}$ and a corresponding tunneling exponent computed from the integral over the classically forbidden region.

However, this description is only valid as long as the nucleating configuration remains separated from the collapsing matter. In the full physical setup, the bubble will inevitably encounter the infalling matter shell after a finite proper time. In our idealized thin-shell description, the collapsing matter separates an exterior Schwarzschild region from an interior Minkowski region, with the matter itself localized on a negligibly thin timelike hypersurface.

\begin{figure}
    \centering
    \def\svgwidth{0.48\linewidth}
    \begingroup%
  \makeatletter%
  \providecommand\color[2][]{%
    \errmessage{(Inkscape) Color is used for the text in Inkscape, but the package 'color.sty' is not loaded}%
    \renewcommand\color[2][]{}%
  }%
  \providecommand\transparent[1]{%
    \errmessage{(Inkscape) Transparency is used (non-zero) for the text in Inkscape, but the package 'transparent.sty' is not loaded}%
    \renewcommand\transparent[1]{}%
  }%
  \providecommand\rotatebox[2]{#2}%
  \newcommand*\fsize{\dimexpr\f@size pt\relax}%
  \newcommand*\lineheight[1]{\fontsize{\fsize}{#1\fsize}\selectfont}%
  \ifx\svgwidth\undefined%
    \setlength{\unitlength}{281.75247313bp}%
    \ifx\svgscale\undefined%
      \relax%
    \else%
      \setlength{\unitlength}{\unitlength * \real{\svgscale}}%
    \fi%
  \else%
    \setlength{\unitlength}{\svgwidth}%
  \fi%
  \global\let\svgwidth\undefined%
  \global\let\svgscale\undefined%
  \makeatother%
  \begin{picture}(1,0.98484752)%
    \lineheight{1}%
    \setlength\tabcolsep{0pt}%
    \put(0,0){\includegraphics[width=\unitlength,page=1]{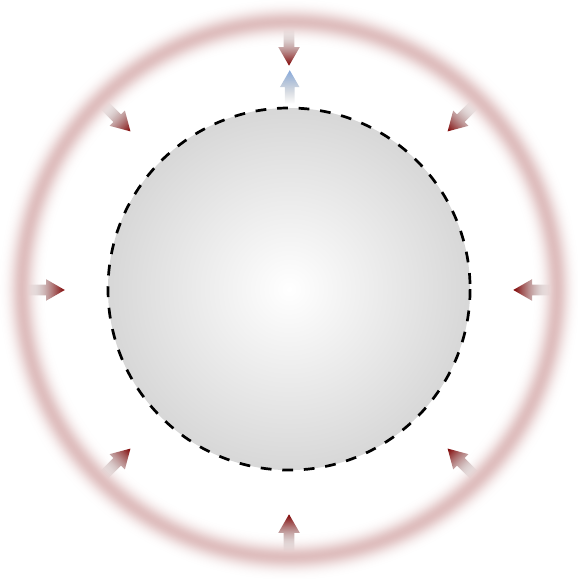}}%
    \put(0.91936261,0.89193947){\color[rgb]{0,0,0}\makebox(0,0)[t]{\smash{\begin{tabular}[t]{c}collapsing\\matter\end{tabular}}}}%
    \put(0.49242389,0.5009487){\color[rgb]{0,0,0}\makebox(0,0)[t]{\smash{\begin{tabular}[t]{c}tunneling\\black shell\end{tabular}}}}%
    \put(0.71074279,0.02523871){\color[rgb]{0.10196078,0.10196078,0.10196078}\makebox(0,0)[lt]{\lineheight{1.25}\smash{\begin{tabular}[t]{l}Schwarzschild\end{tabular}}}}%
    \put(0.56524724,0.14262741){\color[rgb]{0.10196078,0.10196078,0.10196078}\makebox(0,0)[lt]{\lineheight{1.25}\smash{\begin{tabular}[t]{l}Mink$_4$\end{tabular}}}}%
    \put(0.44124526,0.28316312){\color[rgb]{0.10196078,0.10196078,0.10196078}\makebox(0,0)[lt]{\lineheight{1.25}\smash{\begin{tabular}[t]{l}AdS$_4$\end{tabular}}}}%
    \put(0,0){\includegraphics[width=\unitlength,page=2]{blackshell.pdf}}%
  \end{picture}%
\endgroup%

    \caption{Schematic of the collapse geometry for the tunneling configuration, showing the collapsing matter (red) and the tunneling black shell (grey dotted). The AdS, Minkowski, and Schwarzschild regions are indicated. The corresponding effective potential description and its truncation at the matter shell are shown in \cref{fig:potential}. \label{fig:blackshell}}
\end{figure}

From the perspective of the nucleating brane, the spacetime is therefore piecewise composed of an inner AdS–Minkowski junction, governed by the black shell dynamics discussed above, and an outer Minkowski–Schwarzschild junction associated with the collapsing matter shell. This is shown schematically in \cref{fig:blackshell}. When the nucleating configuration reaches the matter shell, the effective potential governing the radial motion ceases to be given by the flat-space expression and is replaced by the full junction dynamics across the matter shell (\cref{fig:potential}).

At this point the tunneling trajectory terminates: the potential is cut off at the radius $R$ where the nucleating shell meets the collapsing matter shell, as shown in \cref{fig:potential-a}.\footnote{We need $r_0>R$ but the qualitative behavior does not depend much on the exact value. If we want the value to be universal and the same for black holes of all sizes, we need $r_0$ to be cosmologically large.} Physically, this corresponds to the black shell “capturing” the infalling matter, with the matter degrees of freedom becoming reprocessed into the shell degrees of freedom. In particular, the collapsing matter is absorbed into the black shell as additional excitations, effectively contributing to the radiation component living on the brane. The tunneling probability is therefore obtained by truncating the Euclidean action at this point, giving
\begin{equation}
    B = 2\int_0^{r=R} |p_{\textrm{\scshape e}} | dr
    \approx \frac{\pi R^2}{2G_4}=\frac{A}{8G_4}\,,
\end{equation}
where we have assumed that $r_0 \gg R$.

\subsection{Enhancing black shell tunneling}

Let us now apply the CDT mechanism we discussed in section 2 to the black shell. For simplicity, we consider the limit of large $r_0 \gg R$
where the potential (defined by evaluating $H$ at $p=0$) becomes
\begin{equation}
    V(r) \approx \frac{r}{G_4}\,,
\end{equation}
and the tunneling amplitude is governed by $p_{\textrm{\scshape e}}=πr/(2 G_4)$. Since the Hamiltonian is not quadratic in $p$ (with a constant mass parameter), the tunneling of the black shell cannot be modeled using a standard Schrödinger equation, as studied in \cref{sec:enhancementsection}. The generalization to the $H(p)$ of the black shell should in principle be straightforward, but in this paper we will only sketch the analysis at a qualitative level.

We assume the existence of large number of degrees of freedom associated with the black shell, which are all coupled to each other. As discussed earlier, this is the structure that makes it possible for the shell to effectively absorb any matter that is dropped onto it. The entropy of the captured mass increases due to the large number of degrees of freedom, rendering the system effectively black. Similarly, as in the case of CDT, we need the interactions to displace the energy eigenvalues such that the effective Euclidean momentum of the leading channels goes to zero. Applying this to the black shell, we expect such a shift in the energy eigenvalue to correspond to a shift in the tension $\sigma_2$ such that $H(p)=0$ at $p=0$. We find that
\begin{equation}
    V(r)=\frac{r}{G_4} \sim 2\sigma g \sqrt{N} \, ,
\end{equation}
assuming a critical tension. For the black shell the number of degrees of freedom are $N\sim n^2$, where $r\sim n l_4$. From this it follows that $\sqrt{g^2\sigma^2/N} \sim 1/l_4 \ll M$. This suggests that the mean fluctuation in the energies of the degrees of freedom of the shell are much smaller than the measured mass of the black shell.

It is important to have the correct picture of what tunneling actually amounts to. It is not that the system suddenly disappears on one side of the barrier and then magically reappears on the other side. It actually passes physically through the barrier along a classically forbidden path. That it is classically forbidden does not mean that it is impossible. The path has still has a finite weight in the path integral, and the use of Euclidean time is a useful trick to take its contribution into account. One should therefore think of the black shell starting at zero radius and physically expanding through the forbidden region until it turns classical at the critical radius.

Let us now imagine a more realistic situation where the matter undergoing collapse is distributed in the bulk. In the simplified example above, all the matter sat at the end of the barrier. More generally, if the matter is distributed over some range in radius, the shell will plow through and interact with matter as it expands under the barrier. The barrier will end if and when you reach the Buchdahl radius of the mass inside of the radius, as indicated in \cref{fig:potential-b}.  All of this matter has been captured by shell. If the Buchdahl radius is never reached, there will be no tunneling and no black shell will (and needs to) be formed. In order for the black shell to look really black, it needs to absorb any matter that is dropped onto it. Its energy will be redistributed among the shell degrees of freedom, but more or less nothing will be emitted back out. The only emission is thermal, and of the order of the Hawking radiation. This efficient absorption is active also under the barrier. As a result, the shell will be in an excited state when it appears on the other side of the barrier carrying all the energy of the matter it has passed through. The interior region has been completely swept clean.

We have presented a qualitative picture of an AdS black shell nucleating via tunneling through a classically forbidden region. A macroscopic black shell has an enormous number of states, more than enough for chaos dominated tunneling to take place. In future work, we hope to carry out a detailed implementation of CDT in this system, taking into account the full dynamics of the black shell as well as considering arbitrary infalling clouds of matter.

\section{On locality}\label{sec:localitysection}

In finding alternatives to black holes it is often argued that it would involve non-local processes at the scale of the Schwarzschild radius. Let us see how this works out for our model.
If the mass is large, the density of a contracting cloud of matter can be extremely low when the event horizon suddenly forms. There is nothing special that happens locally. Another dramatic example is when two black holes collide. An observer in between the two black holes, but outside of both event horizons, will suddenly be inside of the common event horizon. There is a sudden flip from two separate to one common event horizon when the black holes come sufficiently close.
The event horizons are global features, and there is no contradiction with locality in the case of black holes. The worry is that locality would be difficult to maintain if you replace the event horizons with material structures. Yet, our claim is that the tunneling process outlined above achieves exactly this.

Naively, the sudden appearance of a black shell at roughly the same (but a little bit larger) radius as the corresponding Schwarzschild radius, is a clear cut breakdown of locality. However, since it happens through tunneling, with a probability calculated using standard quantum mechanics this cannot be the case. Physics will be consistent, we just need to figure out how.
Recently, the speed of tunneling has been a subject of experimental research \cite{ramos2020measurement,sharoglazova2025energy}. Not surprisingly, the conclusion is that there is no paradox. Naively, it might seem as if a particle can appear on the other side of a barrier faster than the speed of light, but it is always in such a way that no information can be transmitted. This is due to the fact that the tunneling event in itself is out of control. You cannot influence exactly when the tunneling will take place and use this to send a message. This is similar to how entanglement in the context of EPR cannot be used to send information.

To see how this works for the black shell, let us consider a spherical cloud of matter  collapsing towards its Buchdahl radius. When the cloud approaches the critical radius, a quantum shell starts to grow from the center, under the barrier in the forbidden region, sweeping through the cloud and ending up as a classical shell carrying all the matter. Just as for any tunneling event, described in a classical language, it seems as if the center of the cloud “knows” that the radius is becoming dangerously small, and anticipates that it would gain entropy by deciding to tunnel. The crucial point is that there is still no possibility for an observer at the center to use this for communication with the outside at speeds exceeding  the speed of light. 

The actual time of tunneling is not classically determined. Instead you find a superposition of different possibilities. The unfortunate observer inside of the collapsing cloud will not be able to decohere this superposition through a measurement. The number of degrees of freedom of the macroscopic quantum mechanical shell is much too large. Instead, the observer will get entangled with the shell and end up in a macroscopic superposition distinguished by the different possible times for the start of the tunneling.

A young, freshly formed, black hole has an entanglement entropy with the rest of the universe that is no bigger than of the order of the entropy of the matter that formed the black hole. As the black hole grows old, it gets entangled with the rest of the universe. This can be viewed as the universe slowly measuring the black hole. The same is true for the black shell. The black shell can be viewed as a large macroscopic quantum object, given the relative smallness of its initial entanglement with the rest of the universe.
The actual collapse of the wavefunction of the observer will not happen until the black shell is effectively measured through the emission of Hawking radiation in the distant future. A late time observer could in principle figure out an actual measured value of the time of nucleation and thus determine when the unfortunate observer met their demise.

\section{Discussion}\label{sec:discussionsection}

Black hole mimickers offer a conceptually simple yet compelling way out of the problematic aspects of black holes by not forming horizons or curvature singularities. For the black hole mimickers to be physically relevant, these objects have to form at the end of gravitational collapse, and the argument for such a process has largely relied on the large entropy that they carry. These arguments are heuristic and often based on order of magnitude estimates. 

In this paper, we have identified a concrete mechanism that provides a microscopic underpinning for this picture. Our result can be summarized as follows: if a quantum mechanical system tunneling across a barrier has a large number of bright channels, then the transmission coefficient is no longer exponentially suppressed. By studying a one-dimensional particle, we have explicitly shown that such a large number of bright channels can be arranged by suitably choosing the interactions between the internal states. A particularly interesting class is that of chaotic interactions, which are expected to emerge in macroscopic many-body systems, including black shell configurations. The enhanced tunneling through the opening up of a large number of bright channels via chaotic interactions, which we call chaos dominated tunneling (CDT), legitimizes the possibility of the nucleation of black shells and other black hole mimickers during the gravitational collapse of matter.

Although we have argued that the probability for the formation of black shells is no longer exponentially suppressed, it remains to be shown that the probability for the formation of a black shell outweighs that of a black hole at the end of gravitational collapse. While we defer this computation to future work, we can already sketch the argument. The computation would involve finding a suitable notion of a gravitational collapse transition amplitude, whose saddle-point approximation receives contributions from both black hole and black shell configurations. We have seen that it is the total number of bright channels that controls the nucleation of black shells. This is an order-one fraction of the total number of states of the black shell, $N=e^{S_{\textrm{shell}}}$, where $S_{\textrm{shell}}$ is the  entropy of the black shell. It is natural to expect this number to compete against the total number of black hole microstates, $e^{S_{\textrm{\scshape bh}}}$. Since the black shell entropy drastically exceeds the black hole entropy at a given energy, $S_{\text{shell}} \gg S_{\textrm{\scshape bh}}$, the number of bright channels is expected to overwhelm the contribution from the black hole, making the black shell the more probable end product.

\section*{Acknowledgements} 
We would like to thank Thomas Van Riet for useful discussions. VM is supported by a doctoral grant from The University of Iceland Science Park. Support from  Kungliga Fysiografiska sällskapet i Lund is also acknowledged.

\bibliographystyle{JHEP}
\bibliography{refs}

\end{document}